\documentstyle[aps,prl,epsfig,twocolumn]{revtex}

\begin{document}

\title{Sign Rules for Anisotropic Quantum Spin Systems}

\author{R.F. Bishop,$^1$ D.J.J. Farnell,$^2$ and J.B. Parkinson$^1$}

\address{$^1$Department of Physics, University of Manchester Institute of
  Science and Technology (UMIST), P O Box 88, \\
  Manchester M60 1QD, United Kingdom.}

\address{$^2$Institut f\"ur Theoretische Physik, Universit\"at zu K\"oln, 
  Z\"ulpicher Str., 50937 K\"oln, Germany.}

\date{\today}

\address{~
\parbox{15cm}{\rm
\medskip
We present new and exact ``sign rules'' for various 
spin-$s$ anisotropic spin-lattice models. 
It is shown that, after a simple transformation which 
utilizes these sign rules, the ground-state wave 
function of the transformed Hamiltonian is 
positive-definite.
Using these results exact statements for various 
expectation values of off-diagonal operators 
are presented, and transitions in the behavior of these 
expectation values are observed at particular values 
of the anisotropy. 
Furthermore, the effects of sign rules in variational 
calculations and quantum Monte Carlo calculations are 
considered. They are illustrated by a 
simple variational treatment of a one-dimensional 
anisotropic spin model.
[PACS numbers: 75.10.Jm, 02.70.Lq, 03.65.Fd, 75.50Ee]}}


\maketitle

Exact results for quantum many-body systems are rare. They are, 
therefore, always valuable as yardsticks against which numerical
or semi-analytical methods may be measured. Such exact results
acquire additional value when they can be used as actual input to
improve one or more of the approximate methods. Such input is often
useful, for example, in tailoring a trial or starting many-body 
wave function to preserve certain exact properties. In some instances
they can even be practically essential for the successful implementation
of particular techniques. An example is provided by the fixed-node
quantum Monte Carlo (QMC) method, which provides a means to circumvent 
the infamous ``minus sign problem'' inherent in simulating many-fermion
systems, but at the cost of requiring good (and, ideally, exact)
information on the nodal surface of the many-fermion wave function. 
In this paper we provide several exact results of such ``sign rules''
(which exactly define the nodal surfaces) for some general 
classes of anisotropic spin-lattice models. These models 
are of considerable interest both in their own right and 
as models of real (low-dimensional) magnetic materials.

In 1955 Marshall\cite{Marshall} used a variational 
method to study the isotropic spin-half Heisenberg 
antiferromagnet (HAF) specified by the Hamiltonian
\begin{equation}
H=J \sum_{\langle i,j \rangle} {\mathbf s}_i \cdot {\mathbf s}_{j} 
\label{Heis} ~~ ,
\label{eq1}
\end{equation}
where the sum on $\langle i,j \rangle$ counts each nearest-neighbor
pair once and once only, and $J$ is positive. 
Furthermore, the lattice was assumed to be \textit{bipartite} 
(i.e., the lattice can be divided into two sublattices 
($A$,$B$) such that 
all nearest neighbors of a site on one sublattice lie on the other 
and \textit{vice versa}). The total number of atoms was $N$ 
and the spatial dimensionality was not restricted. 
In the course of this paper it was proven that the 
exact ground-state wave function can be written as
\begin{equation}
|\Psi\rangle = \sum_I c_I |I\rangle  ~~ , 
\label{eq2} 
\end{equation}
where $\{|I\rangle\}$ are the usual basis states in the Ising 
representation. The coefficients $\{c_I\}$ were proven to have the 
property that, 
\begin{equation}
c_I=(-1)^{\phi(I)}a_I ~~ ,
\label{eq3}
\end{equation}
where the coefficients $\{a_I\}$ are all positive real numbers or 
zero and $\phi(I)$ is the eigenvalue, with respect to 
the corresponding eigenstate $|I\rangle$,  
of the operator,
\begin{equation}
\phi=n_A ~~,
\label{eq4} 
\end{equation}
which counts all of 
the spin-half `up' states on the $A$-sublattice.
Lieb and Mattis\cite{Lieb1} were able to prove that the ground state of 
the HAF was a singlet following the work of Marshall\cite{Marshall}
and Lieb {\it et al.}\cite{Lieb2}. 

Indeed, these results for the HAF model on a bipartite 
lattice are a consequence of a property of 
non-positive matrices. The Hamiltonian of Eq. 
(\ref{eq1}) is transformed in the following manner,
\begin{equation}
H' = e^{i \pi \phi} H e^{-i \pi \phi} ~~ .
\label{eq5}
\end{equation}
It is seen that $H'$ of Eq. (\ref{eq5}) now contains only 
non-positive off-diagonal interactions with respect to the 
Ising basis states. Hence the ground state of the HAF  
corresponds to the eigenstate of largest magnitude eigenvalue 
which is positive-definite via the Perron--Frobenius 
theorem\cite{Perron}. 
Munro\cite{Munro} and Parkinson\cite{Parkinson} were also 
able to extend these results to the spin-one 
biquadratic model on a bipartite lattice in various phases.
They showed that the ground eigenstates are positive-definite 
and also non-degenerate. Finally, Klein\cite{Klein} 
utilized this knowledge afforded by the Perron--Frobenius theorem
for Heisenberg models to prove six theorems relating to ground-state 
features of these models. Amongst these theorems
was one which stated that the ground-state expectation 
value of ${\mathbf s}_i\cdot{\mathbf s}_j$ for these models is
positive or negative depending on whether $i$ and $j$ are on the same
or different sublattices, respectively.

In this article, we present new sign rules for various 
spin-$s$ anisotropic spin systems. The transformation of Eq. 
(\ref{eq5}) for these sign rules is utilized to force 
the off-diagonal interactions of the Hamiltonians to be 
always non-positive. A property of positive or negative
semi-definite matrices is that the eigenvector corresponding 
to the eigenvalue of $H'$ of largest magnitude (i.e., the 
ground-state energy here) contains only non-negative elements,  
and so the ground-state wave function is positive-definite.

The first model that we consider is the {\it XYZ} model, 
which has a Hamiltonian given by
\begin{eqnarray}
  H_{{\it XYZ}} &=& \sum_{[ i,j ]} \biggl \{ J_{i,j}^x ~ s_i^x s_{j}^x +
  J_{i,j}^y ~ s_i^y s_{j}^y + J_{i,j}^z ~ s_i^z s_{j}^z 
  \biggr \} ~ ,  \nonumber \\
  &=&\sum_{[i,j ]}  \biggl \{ a_{i,j} ~ s_i^z s_{j}^z +
   b_{i,j} (s_i^{+} s_{j}^{-} 
  + s_i^{-} s_{j}^{+}) \nonumber \\
  && ~~~~~~~~~+ c_{i,j}  (s_i^{+} s_{j}^{+}
  + s_i^{-} s_{j}^{-}) \biggr \} ~ , \label{eq6}
\end{eqnarray}
where $a_{i,j}=J_{i,j}^{z}$, $b_{i,j}=(J_{i,j}^{x}+J_{i,j}^{y})/4$ 
and $c=(J_{i,j}^{x}-J_{i,j}^{y})/4$. The index $i$ in Eq. (\ref{eq6}) 
runs over all $N$ sites on a bipartite lattice and $j$ runs over all sites 
which are on the opposite sublattice to $i$. The square bracket 
$[ i,j ]$ indicates that each bond is counted once and once only.   

The second such model that we shall consider is the single-ion 
anisotropy (SIA) model, given by
\begin{eqnarray}
H_{{\rm SIA}} &=& D ~ \sum_i (s_i^x)^2 + 
  \sum_{[ i,j ]} J_{i,j} ~ {\mathbf s}_i \cdot
  {\mathbf s}_{j}  ~~ , \nonumber \\
  &=&\sum_{[ i,j ]} J_{i,j} \biggl \{ s_i^z s_{j}^z +
  \frac 12 (s_i^{+} s_{j}^{-} + s_i^{-} s_{j}^{+}) 
  \biggr \} \nonumber \\
  &+& \frac D4 ~ \sum_i \biggl \{ (s_i^{+})^2 +  (s_i^{-})^2 
  + 2(s(s+1)-(s_i^z)^2) \biggr \} ~. 
\label{eq7}
\end{eqnarray}
Again, the index $i$ in Eq. (\ref{eq7}) runs over all $N$ sites on a 
bipartite lattice and $j$ runs over all sites which are on the 
opposite sublattice to $i$. 

Finally, we also consider the transverse Ising model (TIM), 
which has a Hamiltonian given by
\begin{equation}
H_{{\rm TIM}} = \sum_{[ k_1,k_2 ]} \Omega_{k_1,k_2} ~
s_{k_1}^z s_{k_2}^z  ~ + \frac{\lambda}{2} 
\sum_{i} ~ ( s_i^+ + s_i^- ) ~~ .
\label{eq8}
\end{equation}
For the TIM, we place no restriction on the lattice type and 
$k_1$ and $k_1$ are allowed to run over all lattice sites with
$k_1\ne k_2$. Furthermore, note that
adding extra diagonal terms in Eqs. (\ref{eq6}-\ref{eq8}) 
(such as an external magnetic field or crystal field anisotropy 
in the $z$-direction), or letting $i$ and $j$ in Eqs. (\ref{eq6}) 
and (\ref{eq7}) run over the same sublattice but, in this case, 
explicitly restricting $b_{i,j} \le 0$ in Eq. (\ref{eq6}) 
and $J_{i,j} \le 0$ in Eq. (\ref{eq7}), do not change the 
following sign rules. 

We now define the following operators,
\begin{eqnarray}
m_{A}=\sum_{l_A} s_{l_A}^z ~~ ; ~~
m=\sum_{l} s_l^z ~~, 
\label{eq9}
\end{eqnarray}
where $l_A$ runs over all $N/2$ $A$-sublattice sites and $l$
runs over all $N$ sites on both sublattices. Note that for eigenstate
$|I\rangle=\bigotimes_{l=1}^N  |s_l, m_l\rangle$, where
$s_l^z |s_l, m_l\rangle = m_l |s_l, m_l\rangle$,
$m$ and $m_A$ have eigenvalues denoted by $m(I)$ 
and $m_A(I)$, respectively. Note 
that the following sign rules are fully defined by the 
form of the operator $\phi$ which, in turn, gives the 
set of eigenvalues $\{ \phi(I) \}$. Furthermore, these 
eigenvalues determine the signs of the $\{ c_I \}$ coefficients 
via Eq. (\ref{eq3}) for the expansion of the ground-state wave 
functions, Eq. (\ref{eq2}), of the anisotropic Hamiltonians 
of Eqs. (\ref{eq6}-\ref{eq8}). The sign rules are thus 
defined by,
\begin{equation}
\left .
\begin{array}{ccl@{~~;~~}c@{~~~~}} 
\phi & = & 1               &{\rm rule ~ (A)} \\
\phi & = & m_A             &{\rm rule ~ (B)} \\
\phi & = & m/2             &{\rm rule ~ (C)} \\
\phi & = & m_A+m/2         &{\rm rule ~ (D)} \\
\phi & = & m               &{\rm rule ~ (E)} \\ 
\end{array}
\right \} \label{eq10} 
\end{equation}
Furthermore, note that rule (B) is a reformulation, for general 
spin quantum number, of the Marshall-Peierls sign rule for the spin-half 
HAF given by Eqs. (\ref{eq3}-\ref{eq4}). It is found that the 
first four sign rules apply to the {\it XYZ} and SIA models 
in the following regimes, 
\begin{equation}
\left .
\begin{array}{c@{~~~~}c@{~~~~}c@{~~~~}} 
{\rm Rule}& {\rm {\it XYZ}}~{\rm model} & {\rm SIA}~{\rm model} \\ 
{\rm (A)} & b_{i,j} \le 0, ~c_{i,j} \le 0 
& D \le 0, ~J_{i,j} \le 0 \\ 
{\rm (B)} & b_{i,j} > 0,   ~c_{i,j} \ge 0 
& D \le 0, ~J_{i,j} > 0   \\
{\rm (C)} & b_{i,j} \le 0, ~c_{i,j}   > 0 
& D > 0,   ~J_{i,j} \le 0 \\
{\rm (D)} & b_{i,j} > 0,  ~c_{i,j}   < 0 
& D > 0,   ~J_{i,j} > 0   \\
\end{array}
\right \} \label{eq11}
\end{equation}
for all values of $i$ and $j$ on opposite sublattices. 
(Note again that we do not 
allow the existence of any ``frustrated'' interactions in 
Eqs. (\ref{eq6}) or (\ref{eq7}).) For the TIM, 
it is found that rule (A) applies when $\lambda \le 0$, 
and rule (E) applies when $\lambda > 0$ for 
all signs and strengths of the coefficients $\Omega_{k_1,k_2}$ 
of the Ising interaction in Eq. (\ref{eq8}). 

We once again note that the off-diagonal interactions of $H'$ 
of Eq. (\ref{eq5}) with respect to a complete set of Ising states 
are now always non-positive. Hence the ground-state
eigenvector of $H'$ contains only non-negative elements
and so may be written as $|\Psi'\rangle =\sum_I a_I |I\rangle$,
with $a_i \ge 0 ~ \forall ~ I$.

It is also possible to prove exact relations\cite{Klein} (in some cases) 
for expectation values of various off-diagonal operators. 
In order to illustrate this we define an {\it XXZ}
model for a bipartite lattice in which we set $J_{i,j}^{x} 
=J_{i,j}^y$ and $J_{i,j}^z=1$ in Eq. (\ref{eq6}). Hence we see from the 
table above that the sign rules are of type (A) for $J_{i,j}^x \le 0$ 
and of type (B) for $J_{i,j}^x>0$. We define the spin-spin correlation 
function in the $x$-direction, given by
\begin{eqnarray}
G^{xx}(r) & = & \frac {\langle \Psi | \sum_l s_l^x s_{l+r}^x | \Psi \rangle}
{N\langle \Psi | \Psi \rangle} \label{annex3} ~~ , \label{eq12}
\end{eqnarray}
and the localized magnetization in the $x$-direction to be,
\begin{eqnarray}
M^{x}(l)  &=&  \frac {\langle \Psi | s_l^x | \Psi \rangle} 
{\langle \Psi | \Psi \rangle} ~~.
\label{eq13}
\end{eqnarray}
For the {\it XXZ} model with $J_{i,j}^x>0$, it can be shown that 
if $r$ in Eq. (\ref{eq12}) connects sites on opposite sublattices 
then $G^{xx}(r) \leq 0$ and if $r$ connects sites on the same 
sublattice then $G^{xx}(r) \geq 0$. Similarly for $J_{i,j}^x>0$, 
it can be shown that the sign of $M^x(l)$ when $l$ is a site on 
one particular sublattice must be the opposite to the sign of 
$M^x(l)$ when $l$ is a site on the other sublattice. 
For $J_{i,j}^x<0$, one can show that $G^{xx}(r) 
\geq 0 ~\forall ~r$ and that $M^{x}(l) \geq 0 ~\forall ~l$. 
For the transverse Ising model, it is found for any value 
of $\lambda$ that $G^{xx}(r) \geq 0 ~\forall ~r$. By contrast, 
it is found that $M^{x}(l) \leq 0 ~\forall ~l$ for $\lambda > 0$, 
and that $M^{x}(l) \geq 0 ~\forall ~l$ for $\lambda < 0$. 
We note that other such relations exist for the models 
presented here, although they are not discussed further in
this article.

In order to illustrate the usefulness of the sign rules given here, 
we now perform a simple variational calculation for an anisotropic
spin system which utilizes some of these sign rules. The system in
question is a spin-half one-dimensional model\cite{Bishop1}
in which the anisotropy is in the $x$-direction such that 
$J_{i,j}^x  \equiv -\Delta$,  $J_{i,j}^y=1$ , and $J_{i,j}^z=-1$ 
in Eq. (\ref{eq6}) and $i$ and $j$ are always nearest neighbors.
Hence, from Eq. (\ref{eq11}) the sign rules are of type (A) 
for $ \Delta \ge 1$, of type (D) for $-1 < \Delta < 1$, and of 
type (B) for $\Delta \le -1$. The ground-state energy of this
system using Eq. (\ref{eq2}) may be written as,
\begin{eqnarray}
E &=& \frac{ \sum_{I_1,I_2} c_{I_1}^* c_{I_2} 
  \langle I_1 | H | I_2 \rangle} {\sum_{I'} |c_{I'}|^2} ~~ .
\label{eq14}
\end{eqnarray}

\begin{figure}
\epsfxsize=7cm
\epsffile{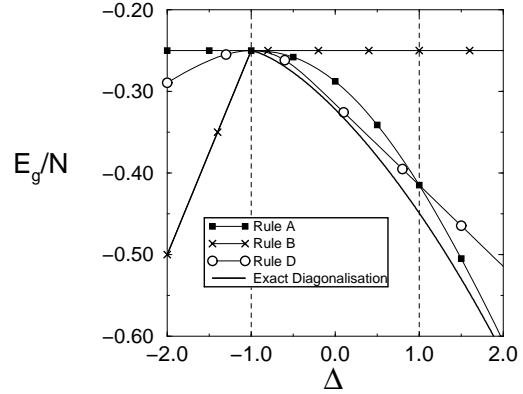}
\caption{Results for the ground-state energy of an anisotropic 
spin-half quantum spin chain (described in text) of length $N=12$. 
The variational calculation utilizes the Jastrow Ansatz of Eq. (\ref{eq15})
in which the sign rules (A), (B), and (D) are used in separate 
calculations and these results are compared to those
of an exact diagonalization of this system.}
\label{fig1}
\end{figure}

In order to treat this model variationally we now employ
a Jastrow Ansatz, given by
\begin{equation}
c_I  =  (-1)^{\phi(I)} \langle I |  \prod_{i<j} 
\biggl [ 1 + f(i,j)  (P_{i}^{\uparrow} P_{j}^{\downarrow} + 
P_{i}^{\downarrow} P_{j}^{\uparrow} ) \biggr ] |I\rangle 
\label{eq15}
\end{equation}
where the $P^{\uparrow}$ and $P^{\downarrow}$ are the usual
projection operators of the spin-half `up' and `down' states 
respectively. The simplest approximation is to set the value of 
$f(i,j)$ to be $f_1$ (a scalar variable) 
if $i$ and $j$ are nearest neighbors, and to be zero otherwise. 
The expectation value of Eq. (\ref{eq14}) may now be evaluated
directly, although we explicitly restrict the sums over all states 
in Eq. (\ref{eq14}) to those states in which $m(I)$ is 
an even number in order to reflect
the symmetries of the Hamiltonian\cite{Bishop1}. The
variational ground-state energy is minimized with respect 
to $f_1$ at each value of $\Delta$, and the different 
sign rules for this system are utilized in separate calculations.
We find that the best results for this variational Ansatz 
(for chains of length $N \le 16$) are those which
utilize the correct sign rule in each regime,
and Fig. \ref{fig1} illustrates these results compared to 
those of an exact diagonalization calculation 
for a chain of length $N=12$. (The accuracy of the variational
calculations with $N=12$ compared to extrapolation in 
the limit $N \rightarrow \infty$ via Pad\'e approximants 
is estimated to be within about 4$\%$ for all $\Delta$.)

This result is clearly true for any case in which one can
prove an exact sign rule (e.g., those defined
by Eqs. (\ref{eq3}) and (\ref{eq10})) 
for a given model because any other choice would 
always contribute at least one non-zero positive contribution
to the expectation energy from one of the `off-diagonal' contributions 
(i.e., $I_1 \ne I_2$) in Eq. (\ref{eq14}). 
We may therefore infer that the best possible variational Ansatz
for a given model must always utilize the correct sign rule in 
the correct regime. 
However, a direct evaluation of the sums in Eq. (\ref{eq14})
scales exponentially with increasing lattice size $N$, 
and so one must evaluate these summations using Monte 
Carlo (MC) techniques. In order to 
explain how this is achieved, we define a {\it probability 
distribution} for the Ising states $\{ |I\rangle \}$, given by 
\begin{equation}
P(I)  = \frac {|c_{I}|^2}{\sum_{I'} |c_{I'}|^2} ~~ ,
\label{eq16}
\end{equation}
and the {\it local energy} of these states, given by
\begin{equation}
E_L(I) = \sum_{I_1} \frac {c_{I_1}}{c_{I}} 
\langle I | H | I_1 \rangle ~~ .
\label{eq17}
\end{equation}
Equation (\ref{eq14}) may be equivalently written as,
\begin{equation}
E  = \sum_{I} ~ P(I) ~ E_L(I) ~~ ,
\label{eq18}
\end{equation}
where the sums over $I_1$ in Eq. (\ref{eq17}) and $I$ in Eq. 
(\ref{eq18}) run over all, or possibly some 
symmetry-constrained subset (e.g., the subset 
of Ising basis states $|I\rangle$ for which $m(I)$ is even),
of the full set of Ising basis states. 
The MC procedure approximates the summation over $I$ 
by performing a directed random walk with respect 
to $|I\rangle$ based on the probability distribution $P(I)$. At each 
point of the MC `summation' procedure one `steps' from one state
$|I\rangle$ to another $|I'\rangle$ (where $|I'\rangle$ is 
one of a number of states accessible to $|I\rangle$ via the
off-diagonal elements of $H$) with a given probability
which is dependent on $P(I)$ and $P(I')$. Hence, it is
possible to `cover' all possible Ising states given 
enough run-time, although one samples the most 
important states the most often. Hence, in this 
manner an accurate approximation (to within 
statistical error) to the sum in Eq. (\ref{eq18}) 
is built up.
Also, the exponentially increasing problem with 
lattice size $N$ is reduced to a problem that scales 
linearly with both $N$ and 
the number of MC moves in a particular run.  
Hence, systems of larger lattice size may be treated.

We note, however, that the sign rule {\it does not} 
affect the random walk which simulates the 
summation over $|I\rangle$ in Eq. (\ref{eq18}) in such a 
variational MC calculation because 
the probability distribution of Eq. (\ref{eq16}) 
is proportional to $|c_I|^2$. By contrast, we may 
see from the expression for the local energy 
of Eq. (\ref{eq17}) that each and every `off-diagonal'
contribution (i.e., $I \ne I_1$) to the local energy 
can only be ensured {\it always} to be negative 
when the correct sign rule in the correct regime is used. 
The average variational ground-state energy, evaluated
using the MC procedure outlined above, is the average 
of the local energies throughout the lifetime
of the run. Hence the average variational MC 
ground-state energy can only ever be ensured to 
be lowest (to within statistical error) when 
the correct sign rule is utilized within the variational 
Ansatz (e.g., in the Jastrow Ansatz of Eq. (\ref{eq15})). 

Let us finally turn our attention from variational MC
estimates to full stochastic simulations of finite 
lattices by QMC techniques. All QMC methods at zero
temperature basically project the exact ground state
$|\Psi \rangle$ of a given many-body Hamiltonian $H$
out of an initial trial state $|\Phi \rangle$ not 
orthogonal to $|\Psi \rangle$, by repeated applications
of some suitable projection operator, $G=G(H)$. This
operation can be formally expressed as a path integral
over many-particle trajectories in configuration space.
Furthermore, it can thus be represented by a stochastic 
process, which may itself be simulated computationally
by the random walks of a sampled set of independent 
``walkers'' through the configuration space. The main
limitation to the applicability of such QMC techniques
arises from the positivity requirement,
\begin{equation}
p(c,c') \equiv \langle c | G | c' \rangle 
\frac {\Phi(c')}{\Phi(c)} > 0 ~ ,
\label{eq19}
\end{equation}
on the probability $p(c,c')$ for a walker to make a 
transition from an initial configuration $|c\rangle$
to another configuration $|c'\rangle$ along a random
walk, where $\Phi(c)$ is the wave function of the 
initial trial state.

For fermionic systems, the minus sign problem associated
with $p(c,c')$ ensures that QMC calculations are much more 
difficult than for bosons. Thus, the computer time for bosonic
simulations to give results of a specified accuracy scales
algebraically with (i.e., as some positive power of) the system
size, $N$. By contrast, when the wave function being sampled
is not positive-definite, and hence cannot itself be regarded
as a probability distribution, no ``exact'' simulation method
has been discovered that does not scale exponentially with 
$N$. Without some prior knowledge of the nodal surface a
QMC simulation will always eventually relax to the 
corresponding bosonic wave function. For the typical
spin-lattice problem for which no such exact knowledge is
available most QMC attempts to alleviate the minus sign 
problem, such as transient estimation, are infeasible
due to its severity. In these cases a recently proposed
lattice variant \cite{haaf} of the fixed-node QMC (FNQMC)
method \cite{ceperley} seems to provide the only practicable
QMC-based approach.

Generally, results of FNQMC calculations are exact, within
statistical errors limited solely by available computing power,
only when the nodal surfaces of both the trial and exact 
ground-state wave functions coincide. Otherwise they yield
variational upper bounds for the energy. On the other hand,
other ground-state quantities such as the (magnetization or)
staggered magnetization may be poorly estimated even when the 
energy is well estimated. How errors in the nodal surface 
affect such quantities is not well understood. In principle,
however, their effects can be large. In this context the
importance of such sign rules as those presented here becomes
clear, in that they permit a complete circumvention of 
the QMC minus sign problem. Indeed, it remains an open 
and challenging problem as to whether such complete or
even partial sign rules can be found for other highly
correlated lattice systems which also suffer from the
minus sign problem. Of particular interest, for example,
would be to attempt to extend our work to both 
electron-lattice models and frustrated spin-lattice
models.

\end{document}